\documentclass[reprint,amsmath,amssymb,aps,prb,]{revtex4-2}
\usepackage{graphicx}
\usepackage{dcolumn}
\usepackage{bm}
\usepackage{siunitx}
\usepackage{amsmath}
\usepackage{ulem}
\usepackage[english]{babel}

\begin{document}
\preprint{APS/123-QED}

\title{Large magnetoresistance and weak-antilocalization in the nodal-line semimetal VP$_2$}
\author{Chunxiang Wu, Shuijin Chen, Tingyu Zhou, Le Liu, Xin Peng, Jianjian Jia, Xinyu Yu}%
\affiliation{%
 Department of Physics, Zhejiang University, Hangzhou 310027, China
}
\author{Hangdong Wang, Jinhu Yang}
\affiliation{
 Hangzhou Key Laboratory of Quantum Matter, School of Physics, Hangzhou Normal University, Hangzhou 311121, China\\
}%
\author{Jianhua Du}
\affiliation{
 Department of Physics, China Jiliang University, Hangzhou 310018, China
}%
\author{Minghu Fang}
\affiliation{%
 Department of Physics, Zhejiang University, Hangzhou 310027, China
 Collaborative Innovation Center of Advanced Microstructures, Nanjing University, Nanjing 20093, China
}
\date{\today}

\begin{abstract}
After growing successfully high quality VP$_2$ single crystals, we studied systematically their longitudinal $\rho_{xx}(T)$ and Hall resistivity $\rho_{yx}(T)$ at various magnetic fields, combining the electronic band and Fermi surface (FS) calculations.
Band calculations reveal that VP$_2$ is a type-II nodal-line semimetal, evidenced by the Hall resistivity measurements.
It is found that the magnetoresistance (MR) at higher magnetic fields exhibits a linear behavior and does not show any sign of saturation, reaching 170\% at 40 K up to 9 T, which is determined by the intrinsic electronic structure and dominated by the Lorenz force, demonstrated by the resistivity anisotropy measurements and the numerical simulations.
We also found that the existence of small amount magnetic impurities (V$^{4+}$, $S=1/2$, 2.24\%) results in Kondo effect emerging in $\rho_{xx}(T)$, the conductivity at lower magnetic fields exhibits a typical weak anti-localization (WAL) behavior.
These results illustrate that VP$_2$ is a platform to study the electronic transport properties of a topological material containing magnetic impurities.

\end{abstract}

\maketitle
\section{INTRODUCTION}
In recent years, topological materials have attracted much research interest due to their novel electrical transport properties, such as extreme magnetoresistance (XMR), ultrahigh mobility, chiral anomaly, anomalous Hall effect, and weak antilocalization (WAL) effects \cite{doi:10.1126/science.1187485,Shekhar2015ExtremelyLM,Ali2014LargeNM,Feng2017ExperimentalRO,PhysRevB.88.104412,PhysRevX.5.031023,PhysRevB.95.195113,PhysRevB.111.195115,adfm.202316775,Zhou2020,PhysRevLett.122.196603}.
These properties originate from their unique electronic structures—including special bulk bands, surface states, or specific Fermi surface geometries—such as Dirac semimetals, Weyl semimetals, and nodal-line semimetals, corresponding to different types of band crossing.
In the nodal-line semimetal, a particularly intriguing class, the valence and conduction bands cross along open lines or closed loops in the Brillouin zone with a nontrivial $\pi$ Berry phase, and topologically protected drumhead surface states \cite{Yang31122022}.

VP$_2$ crystallizes in a monoclinic structure with group C2/m (No.12), as shown in Fig. 1(a), which has the same structure with the transition metal dipnictides (TMDPs) XPn$_2$ (X = Nb, Ta; Pn = P, As, Sb) \cite{PhysRevB.100.205118}, a class of topological materials evidenced by many experimental and theoretical studies \cite{Junjian,Luo2016AnomalousES,Liu2017AnisotropicMI,Li2018,PhysRevB.94.041103}. All these compounds exhibit high mobilities and extremely large positive magnetoresistance (XMR) \cite{PhysRevB.94.041103} due to their unique electronic structure. 
Compared to the 4$d$/5$d$ electrons in NbP$_2$/TaP$_2$ \cite{PhysRevB.93.195106}, the more localization of the 3$d$ electrons in VP$_2$ may introduce a stronger electronic correlation, or result in an additional magnetic scattering of the carriers, as observed by us in VAs$_2$ \cite{Shuijin.Chen:17202}. 
Thus, it is interesting to study the magnetic transport properties of VP$_2$ for understanding the Dirac/Weyl Fermions transport behavior in the systems with magnetic impurities.

In this article, we successfully grew VP$_2$ crystals and measured its longitudinal resistivity, $\rho_{xx}(T,B)$, Hall resistivity, $\rho_{yx}(T,B)$, and magnetic susceptibilities, $\chi(T)$, as a function of magnetic field ($B$) and temperature ($T$), as well as calculated its electronic band structure and Fermi surface (FS). 
Band structure calculations reveal that VP$_2$ is a type-II nodal-line semimetal, with coexistence of electron and hole pockets at FS.
It is found that $\rho_{xx}(T,B)$ exhibits a minimum at 46 K, which is attributed to the scattering of magnetic impurities (i.e., Kondo effect) V$^{4+}$ ($S=1/2$) evidenced by both the fitting of $\rho_{xx}(T)$ data and $\chi(T)$ measurements. 
The magnetoresistance (MR) at higher magnetic fields exhibits a linear behavior and does not show any sign of saturation, which is determined by the intrinsic electronic structure, demonstrated by the resistivity anisotropy measurements and the numerical simulations. 
However, the conductivity at lower magnetic fields exhibits a typical weak anti-localization (WAL) behavior, implying that the scattering of magnetic impurities to the carriers is dominant.

\section{Experimental Methods}

\begin{figure*}[ht]
	\includegraphics[width=1\textwidth]{fig_1.jpg}
	\caption{(a) Crystal structure of VP$_2$. (b) XRD pattern of a VP$_2$ polycrystal crystal. Inset: an image of typical single crystals. (c)  XRD pattern of a VP$_2$ single crystal. Inset: EDS data of a cleaved sample surface. (d) The calculated Fermi surface of VP$_2$ (e) The nodal-line in first Brillouin zone without considering SOC. (f) high symmetry points in the reciprocal primitive cell. (g) and (k) Fatband calculated without considering SOC (g) and taking SOC into account (k). The colors of the line represents the projectors for different atom orbits and the thickness of the line represents the weight.}
\end{figure*}

VP$_2$ single crystals were grown by a chemical vapor transport method. High purity V and P powders were mixed in a mole ratio 1 : 2, sealed in an evacuated quartz tube, containing I$_2$ as a transport agent with 10 mg/cm$^3$.
The quartz tube was placed in a tube furnace with a gradient 200$^\circ$C, then heated slowly to 900$^\circ$C at its hot end and sustained for two weeks, finally cooled to room temperature after power being turned off. Needle-like crystals with a typical dimension of $5\times 0.07\times 0.05$ mm$^3$ [see the inset of Fig. 1(b)] were obtained at the cold end. The composition was determined by using the energy dispersion x-ray spectrometer (EDXS) to be V : P = 33.3 : 66.7. The crystal structure was determined using a powder x-ray diffractometer (XRD, Rigaku Gemini A Ultra) by using sample produced by grinding pieces of crystals. It was confirmed that VP$_2$ crystallizes in a monoclinic structure (space group C2/m, No.12). 
The lattice parameters $a$ = 8.463(2) \AA, $b$ = 3.105(5) \AA, $c$ = 7.169(1) \AA  ($\alpha$ = 90$^\circ$, $\beta$ = 119.26(9)$^\circ$, $\gamma$ = 90$^\circ$) were obtained by using the Rietveld refinement to XRD data (the weight profile factor $R_{wp}$ = 7.96$\%$ ,and the goodness-of-fit $\chi^2$ = 2.105), as shown in Fig. 1(b), consistent with the results reported in Ref. \cite{GOLIN1975}. 
Longitudinal resistivity ($\rho_{xx}$), Hall resistivity ($\rho_{yx}$), and magnetization measurements were carried out by using a Quantum Design physical property measurement system (PPMS-9T) and magnetic property measurement system (MPMS-7T), respectively. The band structure calculations were performed using the Vienna $ab initio$ simulation package (VASP) \cite{PhysRevB.54.11169,PhysRevB.59.1758} with a generalized gradient approximation (GGA) of Perdew, Bunke, and Ernzerhof (PBE) \cite{PhysRevLett.77.3865} for the exchange-correlation potential. A cut-off energy of 400 eV and a $25\times 25\times 11$ k-point mesh were used to perform the bulk calculations.
The VASP–WANNIER90 interface was employed to obtain the tight-binding Hamiltonian, utilizing $s$, $p$ and $d$ orbitals of V atoms, as well as $s$ and $p$ orbitals of P atoms \cite{MOSTOFI2008685}. The nodal-line search, Fermi surface (FS) structures and ordinary magnetoresistance calculations were performed by the open-source software WannierTools \cite{WU2017}.

\section{Results and Discussions}

At first, we discuss the results of our electronic band structure and FS calculations that extend the initial prediction of high symmetry line semimetal for VP$_2$ \cite{Nature.566.475}. As shown in Fig. 1(d), there are six different FS pockets: two hole-like surfaces (blue) at the $\mathrm{L}$ and $\mathrm{Y}$ points, and four electron-like (red) surfaces near the $\mathrm{L}$ and $\mathrm{Z}$ points. 
Figures 1(g) and 1(k) show the bulk bands without and with considering spin-orbit coupling (SOC), respectively. It is clear that the bands near the Fermi level ($E_\mathrm{F}$) mainly arise from the $d$ orbits of V atoms, and the valence and conduction bands cross along the $\mathrm{Y}-\mathrm{X}_1, \mathrm{Z}-\mathrm{I}_1, \mathrm{L}-\mathrm{I}$ high symmetry directions. 
Without SOC, in the Brillouin Zones (BZ) there is only one type nodal-line, four nonclosed, spiral lines extending across the BZ around Z point, see Fig. 1(e), different with those (two types nodes) existing in the identical structure VAs$_2$ \cite{Shuijin.Chen:17202}. When SOC is considered, these nodal-lines are gaped [see Fig. 1(k)] to a band inversion along the $\mathrm{Y}-\mathrm{X}_1,\mathrm{Z}-\mathrm{I}_1,\mathrm{L}-\mathrm{I}$ high symmetry directions, driving into a topological crystalline insulator (TCI), as discussed by Wang $et$ $al.$ \cite{PhysRevB.100.205118} for the identical structure TMDPs XPn$_2$. Although the opening of a local band gap between the valence and conduction bands occurs when SOC is included, VP$_2$ preserves its semimetal character with the presence of electron and hole pockets, similar to that in NbAs$_2$ \cite{Li2018,PhysRevB.94.041103}, and VAs$_2$ \cite{Shuijin.Chen:17202}.

\begin{figure}[ht]
	\includegraphics[width=0.85\textwidth]{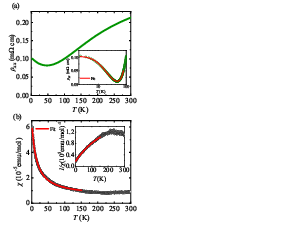}
	\caption{(a) Temperature dependence of resistivity. (b) Temperature dependence of susceptibility. Inset: reciprocal susceptibility as a function of temperature.}
\end{figure}

Second, we focus on the Kondo effect occurring in the longitudinal resistivity for VP$_2$. Figure 2(a) presents the temperature dependence of resistivity, $\rho_{xx}(T)$. With decreasing temperature, the resistivity $\rho_{xx}$ decreases monotonously from $\rho_{xx}$(300 K) = 213.36 \si{\micro\ohm} cm, reaches a minimum at about 46 K, then increases to 100.88 \si{\micro\ohm} cm at 2 K, thus the residual resistivity ratio (RRR) $\rho_{xx}$(300 K)/$\rho_{xx}$(2 K) $\sim$ 2.1, much smaller than that observed in the other transition metal dipnictide crystals, such as VAs$_2$ ($\sim$12) \cite{Shuijin.Chen:17202}, NbAs$_2$ ($\sim$75) \cite{PhysRevB.94.041103}, TaAs$_2$ ($\sim$100) \cite{Recentadvancesintendinopath}. As we know, the system containing magnetic impurities, which usually scatter the conduction electrons though the $s-d$ exchange interactions, exhibits a minimum in resistivity at lower temperature, $i.e$, termed as the Kondo effect \cite{Kondoeffect,PhysRevB.7.3215}. 
As discussed by us for VAs$_2$ \cite{Shuijin.Chen:17202}, we use a simplified Kondo model to analyze, in which the temperature dependent resistivity is given by \cite{PhysRevLett.107.256601}

\begin{equation}
\rho=\rho_0+qT^2+pT^5+\rho_\mathrm{K}
\end{equation}


where $\rho_0$ is the residual resistivity due to sample disorder, the $T^2$ and $T^5$ terms are the electron-electron and electron-phonon scattering contributions, respectively, the $\rho_\mathrm{K}$ is the Kondo effect term, which is given by:

\begin{equation}
\rho_\mathrm{K}=\rho_\mathrm{K}(0 {\rm{K}})(\frac{{T_\mathrm{K}^{\prime}}^2}{T^2+{T_\mathrm{K}^{\prime}}^2})^s
\end{equation}
\begin{equation}
T_\mathrm{K}^{\prime}=\frac{T_\mathrm{K}}{(2^{1/s}-1)^{1/2}}
\end{equation}

where $T_\mathrm{K}$ is the Kondo temperature and  $s$ = 0.225 for a spin 1/2 impurity system according to the numerical renormalization group theory \cite{Costi_1994,PhysRevLett.81.5225}. We used Eq. (1) to fit the resistivity data measured at low temperatures (2 - 80 K), as shown in the inset in Fig. 2(a). It is clear that Eq. (1) can well described the $\rho_{xx}(T)$ data below 80 K and the obtained parameters in Eqs. (1) and (2) from the fitting are listed in Table 1.


\begin{table}[h]
	\caption{Fitting results of low-temperature resistivity $\rho_{xx}(T)$.}	
\begin{ruledtabular}  
	\begin{tabular}{ccc}
$\rho_0$ ($\rm \mu \Omega$-$\rm cm$)&
$q$ ($\rm \mu \Omega$-$\rm cm$-$\rm{K}^{-2}$)&
$p$ ($\rm \mu \Omega$-$\rm cm$-$\rm{K}^{-5}$) \\ \hline

49.1 &
$2.2 \times 10^{-3} $ &
$14.1 \times 10^{-10} $

\\ \hline 

$\rho_\mathrm{K}(0{\rm{K}})$ ($\rm \mu \Omega$-$\rm cm$)&
$T_\mathrm{K}$ ($\rm K$) 

\\	 \hline
51.8    & 
56.1
	\end{tabular}
\label{tab kondo fit}
\end{ruledtabular}
\end{table}

The existence of V$^{4+}$ (S = 1/2) impurities in our VP$_2$ crystals was also confirmed by the magnetic susceptibility measurements. Figure 2(b) shows the temperature dependence of susceptibility, $\mit\chi(T)$, measured at 3 T with a field-cooling (FC) process. 
With decreasing temperature, the susceptibility $\mit\chi$ decreases a little, reaches a minimum at around 220 K, then increases strikingly below 100 K. No magnetic transition was observed in the whole temperature range (2 - 300 K) and the significant increase of $\mit\chi$ in low temperatures was believed to originate from the contribution of V$^{4+}$ ($S$ = 1/2) impurities existing in crystals as the interstitial ions, as observed in VAs$_2$ and VSe$_2$ crystals \cite{Shuijin.Chen:17202,SignaturesoftheKondoeffectinVSe2}. We used the modified Curie-Weiss law $\mit\chi = \mit\chi_0+\frac{C}{T-\mit\theta}$ to fit the data below 150 K, as shown in Fig. 2(b), the Curie constant $C$ = 8.37 ($\pm 0.07$) $\times$ 10$^{-4}$ emu K/mol and the Curie temperature $\theta$ = 13.6 ($\pm0.1$) K, the temperature-independent $\chi_0=5.09$ $(\pm 0.07)\times10^{-6}$ emu/mol were obtained. The nearly linear relationship between $\frac{1}{\mit\chi}$ and $T$ below 150 K [see the inset in Fig. 2(b)] demonstrates the reliability of the fitting. The V$^{4+}$ ($S$ = 1/2) impurity molar fraction was estimated to be of 2.24 ($\pm0.02$)\%, small amount magnetic impurities existing in the crystals.

\begin{figure*}[htbp]
\includegraphics[width=1\textwidth]{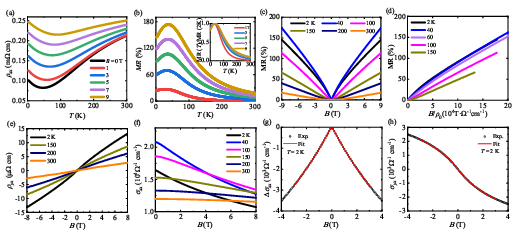}
\caption{(a) Temperature dependence of resistivity of $\mathrm{VP}_2$ single crystal under different magnetic fields. 
(b) MR as a function of temperature under different magnetic fields. Inset: Normalized MR as a function of temperature under different magnetic fields. (c) MR as a function of magnetic field at different temperatures. (d) Fitting of MR using Kohler’s law. (e) Hall resistivity $\rho_{yx}(B)$ as a function of magnetic field at various temperatures. (f) Longitudinal conductivity $\sigma_{xx}(B)$ as a function of magnetic field at different temperatures. (g) Fitting of longitudinal conductivity change $\Delta \sigma_{xx}(B)$ at $T$ = 2 K. (h) Fitting of Hall conductivity $\sigma_{xy}(B)$ at $T$ = 2 K.}
\end{figure*}

\begin{figure}[h]
\includegraphics[width=0.5\textwidth]{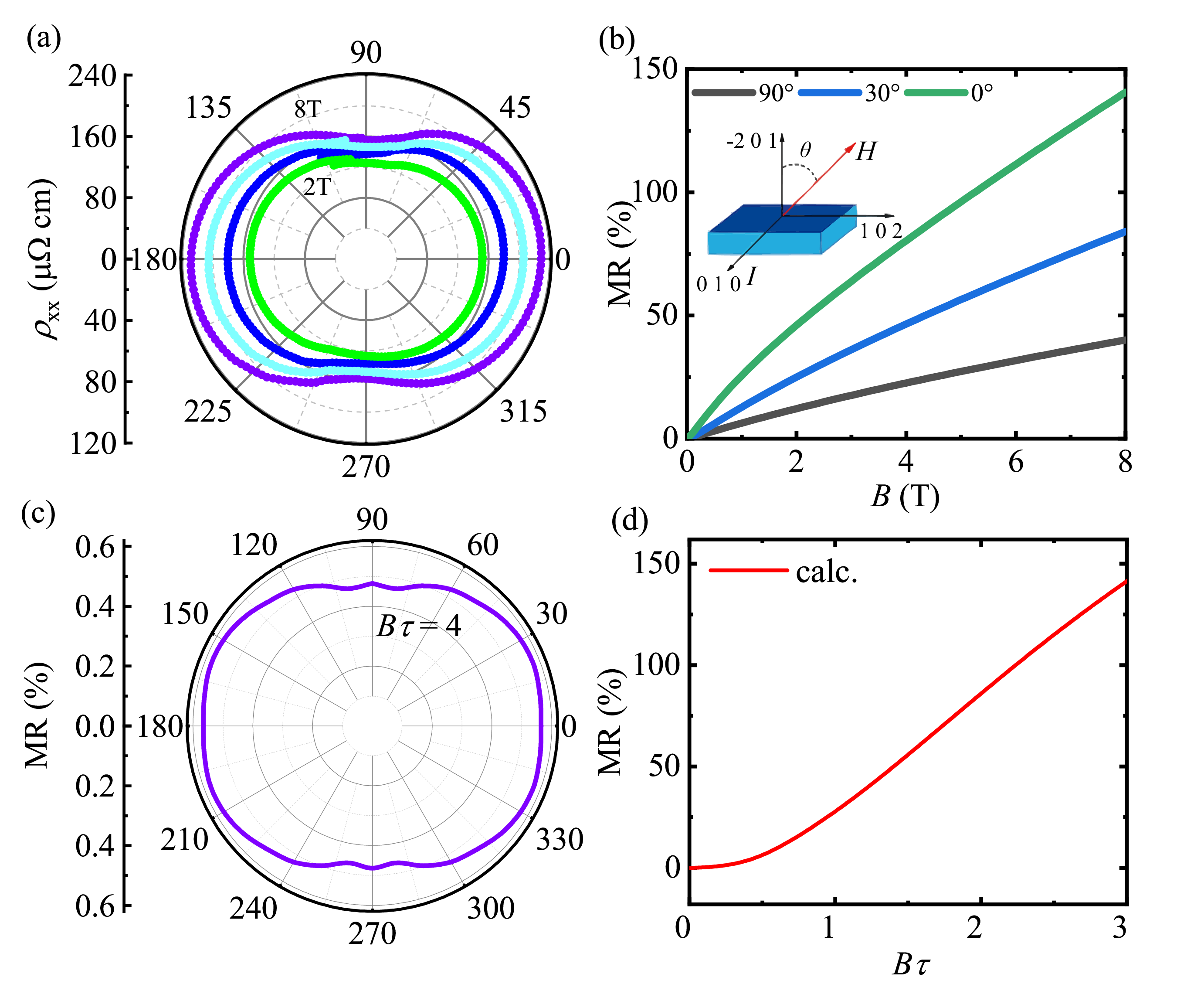}
\caption{(a) Angular dependence of resistivity $\rho_{xx}$ at 2 K for different magnetic fields rotated in the $a-c$ plane. (b) MR as a function of magnetic field magnitude for $\theta = 0^\circ$, $30^\circ$, $60^\circ$, and $90^\circ$ at 10 K. Inset: Schematic of magnetic field orientation. (c) Calculated angular dependence of MR at $B\tau = 4$. (d) Calculated MR for $\theta = 0^\circ$.}
\end{figure}

Third, we discuss the MR occurring in this nodal-line semimetal VP$_2$ with the presence of Kondo scattering of V$^{4+}$ impurities. Figure 3(a) shows the temperature dependence of longitudinal resistivity, $\rho_{xx}(T)$, measured at various magnetic fields $B$, with current $I$ applied in the $b$ axis, and $B \perp b$ axis. 
It is obvious that the resistivity in magnetic field increases significantly in the whole temperature range (2 - 300 K), except that the resistivity measured at $B$ = 1 T increases significantly only in lower temperature ($<$ 150 K) range. 
This behavior is different with that observed in many other nontrivial and trivial topological semimetals \cite{YE2025184934,Shuijin.Chen:17202,Junjian,Peng2025}, in which, an up-turn in $\rho(T)$ curves under applied magnetic fields occurs at low temperatures, $i.e.$ $\rho$ increase with decreasing $T$ and then saturates. Figure 3(b) shows the magnetoresistance (MR) with the conventional definition MR = $\frac{\rho(B)-\rho(0)}{\rho(0)}\times100\%$, as a function of temperature measured at various fields. With increasing temperature, the MR increases at lower temperatures, and reaches a maximum at the temperature where resistivity having a minimum, thus decreases. Even at room temperature (300 K), the MR has a considerable value (17$\%$ at 9 T). The normalized MR by the maximum, shown in the inset of Fig. 3(b), has the same temperature dependence for various fields except for $B$ = 1 T, indicating that the MR mechanism is the same at high magnetic fields in the whole temperature range (2 - 300 K), however, different at low magnetic fields. 
The field dependence of MR at various temperatures is shown in Fig. 3(c). 
MR increases nearly linearly at higher fields without any signature of saturation in the whole temperature range (2 - 300K), but its low-field behavior is temperature-dependent, a dip emerges at high temperatures, while a hump develops at low temperatures.


We tried to fit the MR data using the Kohler's rule \cite{PhysRevB.92.180402},
$\mathrm{MR} \propto \left(\frac{B}{\rho_0}\right)^m$,
as discussed for many trivial/non-trivial semimetals, as well as discussed by us for VAs$_2$ with the same structure \cite{Shuijin.Chen:17202}. 
As shown in Fig. 3(d), it is clear that the Kohler's rule cannot describe the MR data obtained in the whole temperature range (2 – 300 K). Even we fitted the MR data using the extended Kohler's rule \cite{PhysRevX.11.041029}, introducing a thermal factor $n_T$ for the carrier concentration changing with temperature, i.e., $\mathrm{MR} \propto \left(\frac{B}{\rho_0 n_T}\right)^m $.
It was found that the extended Kohler's rule cannot describe the MR data, too, implying that other scattering mechanism may occur in our crystals.

To understand the mechanism of MR at higher magnetic fields, we measured the longitudinal resistivity anisotropy at various magnetic fields. 
Figure 4(a) shows the angular resistivity polar plot $\rho_{xx}(\theta)$ measured at 10 K in $B$ = 2, 4, 6 and 8 T with the current $I$ along the $b$ axis and rotating $B$ in the $a$-$c$ plane [see the inset of Fig. 4(b)]. The $\rho_{xx}(\theta)$ at 2 K exhibits a twofold symmetry, i.e., $\rho_{xx}(\theta)$ =$\rho_{xx}(\theta+\pi)$, which is consistent with the monoclinic structure of VP$_2$. 
We choose $\theta = 0, 30^\circ$, and $90^\circ$ to measure $\rho_{xx}(B)$ at 2 K, and calculate the MR($B$), as shown in Fig. 4(b). As the external magnetic field $B$ is applied along the [-2 0 1] direction ($\theta = 0^\circ$), the MR has a maximum, reaching 140\% at 8 T, while $B \parallel [102]$ ($\theta = 90^\circ$) MR = 40\%. The MR($B$) exhibits a similar linear behavior in the higher magnetic field range for these three directions. 
These results imply that the magnetic transport properties at higher magnetic field is determined by the geometry of FS.

Figure 4(c), (d) shows our numerical simulation results for resistivity anisotropy and the magnetic field dependence of MR, respectively, by combining the FS discussed above with the Boltzmann transport theory approach based on the semiclassical model and the relaxation time approximation by using the open-source software WannierTools \cite{WU2017}, as done by us for VAs$_2$ \cite{Shuijin.Chen:17202}, SiP$_2$ \cite{PhysRevB.102.115145}, RuO$_2$ \cite{Peng2025} and CrSb \cite{PhysRevB.111.144402}.
It is clear that the calculated anisotropy of resistivity for $B$ rotated in the $a$-$c$ plane agree well with the experimental data shown in Fig. 4(a). 
The calculated magnetic field dependence of MR also exhibits a linear behavior at higher magnetic field region, consistent with the experimental results at 2 K for $\theta = 0^\circ$, shown in Fig. 4(b).
However, the calculated MR($B$) in lower magnetic range, a quadratic dependence, exhibits obviously a different behavior with that of the experimental data [Fig. 4(b)], confirming further the existence of another scattering mechanism in lower magnetic fields, as discussed below.

For understanding the MR behavior at lower magnetic fields, we measured the Hall resistivity, $\rho_{yx}(B)$, at various temperatures, as shown in Fig. 3(e), to obtain the magnetoconductivity (MC) $\sigma_{xx}(B)$ and $\sigma_{xy}(B)$.
We calculated the longitudinal conductivity $\sigma_{xx}$ = $\rho_{xx}$/($\rho_{xx}^2$+$\rho_{yx}^2$), and the Hall conductivity $\sigma_{xy}$ = -$\rho_{yx}$/($\rho_{xx}^2$+$\rho_{yx}^2$) by using the original experimental $\rho_{xx}(B)$ and $\rho_{yx}(B)$ data. 
Figure 3(f) shows the obtained $\sigma_{xx}(B)$ , which shows a sharp decrease in low field at low temperatures, remind us that a weak antilocalization (WAL) effect may occur in our VP$_2$ crystals \cite{10.1117/12.2063426}.
WAL is a quantum interference phenomenon in which electrons traversing closed trajectories acquire an additional $\pi$ Berry phase, leading to destructive interference that enhances backscattering and thereby reduces conductivity. When an external magnetic field is applied, it disrupts this phase coherence, resulting in a rapid decrease in $\sigma_{xx}(B)$.
This phenomenon has been widely observed in various topological insulators and semimetals \cite{PhysRevB.95.195113,PhysRevB.111.195115,adfm.202316775,Zhou2020,PhysRevLett.122.196603},  in which electrons around Dirac points or nodal-lines can pick up an extra $\pi$ Berry phase.
In lower magnetic fields, the conductivity change $\Delta \sigma_{xx}(B)=\sigma_{xx}(B)-\sigma_{xx}(0)$ due to WAL can be fitted with the Hikami-Larkin-Nagaoka (HLN) equation \cite{10.1143/PTP.63.707}:

\begin{equation}\begin{aligned}
\Delta \sigma_{xx}(B) =-\frac{\alpha e^{2}}{\pi\hbar}\left[\Psi\left(\frac{1}{2}+\frac{\hbar}{4e{l_{\phi}}^{2}B}\right)-\ln\left(\frac{\hbar}{4e{l_{\phi}}^{2}B}\right)\right]
\end{aligned}\end{equation}

where $\alpha$ is a prefactor, $\hbar$ is the reduced Planck constant, $\Psi$ is the digamma function, $l_\phi$ is the phase coherence length.
As an example, Figure 3(g) presents the HLN fitting at $T$ = 2 K in the magnetic range -3 T $\leq B \leq$ 3 T, yielding $\alpha=1.7 \times 10^9$ and $l_\phi=36$ nm.
Notably, the extracted $\alpha$ value is significantly larger than the typical value of $\alpha \sim 1$ expected for a single coherent conduction channel in conventional topological insulators \cite{10.1143/PTP.63.707}. Such a large $\alpha$ strongly suggests the contribution of multiple conducting channels likely originating from bulk states transport, rather than from two-dimensional surface states.
Chen $et$ $al.$ theoretically analyzed MC of nodal-line semimetals at low fields follows a $-ln(B)$ dependence for the 2D WAL effect and a $\sqrt{B}$ dependence for the 3D WAL effect \cite{PhysRevLett.122.196603}.
Their result is specific to torus-shaped Fermi surfaces while other FS geometries could also show a $-ln(B)$ behavior, as recently observed in the CaAgBi \cite{Sasmal_2020} and CaZn$_2$Sb$_2$ \cite{PhysRevB.111.195115}.
In VP$_2$, the FS is clearly not torus-shaped, so it is reasonable to use $-ln(B)$ to fit its conductivity, as a limiting of Eq. (4) for $B \gg {\hbar} / {4el_\phi^2}$.

Another, we also noticed that the Hall resistivity exhibits a nonlinear behavior at lower temperatures and lower fields, which is a feature of WAL only in multi-carrier systems.
Following the analysis for Bi$_{1 -x}$Sb$_x$ by Duy $et$ $al.$ \cite{PhysRevB.100.125162}, using the modified semiclassical two-carrier model, longitudinal and Hall conductivity can be given by 

\begin{equation}
\begin{aligned}
\sigma_{xx}(B) & =\frac{n_ee\mu_e}{1+\mu_e^2B^2}+\frac{n_he\mu_h}{1+\mu_h^2B^2} \\
 & =\frac{n_he^2D_e}{1+e^2D_e^2B^2}+\frac{n_he^2D_h}{1+e^2D_h^2B^2}
\end{aligned}
\end{equation}
\begin{equation}
\begin{aligned}
\sigma_{xy}(B) & =\left(\frac{n_{e}e\mu_{e}^{2}}{1+\mu_{e}^{2}B^{2}}-\frac{n_{h}e\mu_{h}^{2}}{1+\mu_{h}^{2}B^{2}}\right)B \\
 & =\left(\frac{n_{h}e^{3}D_{e}^{2}}{1+e^{2}D_{e}^{2}B^{2}}-\frac{n_{h}e^{3}D_{h}^{2}}{1+e^{2}D_{h}^{2}B^{2}}\right)B
\end{aligned}
\end{equation}

where $n_e$ and $n_h$ denote the carrier concentrations, and $\mu_e$ and $\mu_h$ denote the mobility of electrons and holes, respectively. 
$D_{e(h)}$ is the diffusion coefficient of electrons (holes), related to the mobility by  $\mu_{e(h)}=eD_{e(h)}$, and is modified by WAL effect.
Applied external magnetic field decreases the value of $D_{h(e)}$, thereby affecting both the longitudinal and Hall conductivity \cite{PhysRevB.100.125162}. 
For convenience, we considered only the correction to the conductivity from $D_h$ when fitting.
We estimate $\Delta D_h$ using the approximate relationship $\Delta D_h \approx ne^2 \Delta \sigma_{xx}$, and then obtain $D_h(B)=\Delta D_h+D_h(0)$ \cite{Salawu2022}.
The fitting results at $T$ = 2 K, for example, are shown in Fig. 3(h), yielding $n_e=5.6 \times 10^{-19}$ cm$^{-3}$, $n_h=7.1 \times 10^{-19}$ cm$^{-3}$, $\mu_e=0.05$ cm$^2$V$^{-1}$s$^{-1}$ and $eD_h(0)=0.11$ cm$^2$V$^{-1}$s$^{-1}$, implying that the non-linearity of Hall resistivity at low field is due to the WAL correction.
The correction of the Hall conductivity by WAL can only be observed in multi-carrier systems, because in a single-carrier material, the Hall coefficient $R_\mathrm{H}=1/ne$ is completely independent of the mobility $\mu$.



Although magnetic impurities are generally expected to suppress WAL through enhanced spin-dependent scattering that destroy time-reversal symmetry, our observation of magnetic impurities (V$^{4+}$, $S=1/2$) coexisting with WAL in VP$_2$ demonstrates that such suppression is not inevitable.
This coexistence is also supported by a recent work on magnetic half-Heusler compounds RPtBi (R = lanthanide), where Chen $et$ $al.$ reported a systematic crossover from WAL to weak localization (WL) by tuning the lanthanide element from Tm ($S=1$) to Tb ($S=3$) \cite{10.1063/1.5143990}.
In those systems, WAL persists in compounds with smaller spin angular momentum $S$ (TmPtBi and ErPtBi), while stronger spin-dependent scattering in DyPtBi and TbPtBi drives the transition to WL.
These results indicate that weak magnetic scattering associated with $S = 1/2$ impurities in VP$_2$ is compatible with WAL.
Our observation thus provides another example that magnetic scattering and WAL can indeed coexist, extending the understanding of quantum interference in magnetically doped systems.\newline

\section{CONCLUSIONS}
In summary, we successfully grew VP$_2$ single crystals by using a chemical vapor transport method, and systematically investigated its magnetotransport properties, combing with the electronic band and FS calculations.
Band calculations reveal that VP$_2$ is a type-II nodal-line semimetal with coexisting of electron and hole pocket evidenced by the Hall resistivity measurements with a multi-band characteristics.
It is found that the MR at higher magnetic field exhibits a linear behavior and does not show any signals of saturation, reaching 170\% at 40 K in 9 T, similar to those observed in many nontrivial/trivial semimetals, which is determined by the intrinsic electronic structure and dominated by the Lorentz force, evidenced by the resistivity anisotropy measurement and the numerical simulations. 
We found that the small amount magnetic impurities (V$^{4+}$, $S=1/2$) result in the Kondo effect emerging in $\rho_{xx}(T)$, the conductivity at lower magnetic fields exhibits typical WAL effect. 
These results demonstrate that VP$_2$ provides a material platform to study the electronic transport properties of topological material containing magnetic impurities.

\section{ACKNOWLEDGMENTS}
This research is supported by the National Key Research and Development Program of China under Grant No. 2022YFA1403202, the National Natural Science Foundation of China under Grant No. 12074335

\bibliography{1.bib}
\end{document}